# Deep learning-based dynamic error correction and uncertainty estimation for digital twin-assisted fringe projection profilometry of rotating gears

zhang sheng Li    jian cheng Qiu    gao xu Wu*

*Jiangxi University of Science and Technology*

## Abstract

This paper presents a deep learning-based method for dynamic gear measurement and uncertainty estimation. A digital twin system based on Unity is used to flexibly generate diverse simulated datasets, making it easy to obtain large amounts of simulation data for convenient verification of network performance, especially when real data is difficult to acquire. The proposed CD-DCU-Net integrates Concrete Dropout for pixel-wise uncertainty estimation and employs dual 3×3 convolutions in the output layer to enhance the spatial continuity of the predictions. During network training, a transfer learning strategy is adopted: the network is first pretrained with only a small amount of simulated data, and then further trained on the target dataset. In addition, a weighted loss function is introduced to better balance phase prediction accuracy and uncertainty estimation. Experimental results show that, compared with the traditional three-step phase-shifting (3-step PS) method, the proposed approach achieves significant improvements in phase prediction, 3D reconstruction accuracy, dynamic error correction, and uncertainty estimation, providing a practical and efficient solution for fringe projection measurement of gears.

*Keywords: Fringe projection; Dynamic Measurement; Deep learning; Digital twin; Gear measurement; Three-step phase-shifting*

## 1 introduction

The gears are manufactured with discontinuous surfaces, repetitive features, and significant height variations, making full-tooth three-dimensional (3D) measurement a technical challenge, especially in dynamic applications where efficiency is of the utmost importance. In recent years, fringe Projection profilometry (FPP) has become one of the promising method due to its advantages of non-contact operation, high precision, and high efficiency, and attracted increasing attention in the field of gear measurement[1, 2]. A typical FPP-gear system contains one projector for active coding and one camera for image capturing, with a common field of view covering the width of the gear teeth. By coding projector's pixel coordinate information as the phase map into pattern grayscale, pixel-level correspondence modulated by gear tooth can be established from the camera's capturing images. The complex 3D profile information of the gear can be then calculated pixel-by-pixel based on triangulation principle[3]. Currently, FPP is more focused on in static measurement such as the manufacturing accuracy, while the measurement requirement of dynamic characteristics of the gear is hardly improved.

To further improve the efficiency and accuracy of FPP in gear measurement, researchers now spare no effort to improve or create new pattern encoding and decoding methods[4]. Mainstream methods can be divided into multi-frame phase decoding methods and single-frame phase decoding methods, with representative methods being phase shifting profilometry (PSP) and Fourier transform profile measurement (FTP) respectively. PSP performs pixel-independent phase decoding based on the time-axis sampling by projecting multi frames, and it's generally ideal for high-precision measurements on complex but static surface[5]. In contrast, FTP uses only one fringe pattern, thus enables fast and real-time measurement. FTP is most advantageous for dynamic scenarios, but it cannot deal with complex surfaces especially discontinuities due to its dependence on regional information[6]. In summary, for the application of rotating gear, both single-frame (represented by FTP) and multi-frame (PSP) approaches exhibit inherent technical limitations. FTP is prone to accuracy loss when dealing with large height variations on tooth geometries, whereas PSP is usually better suited for complex gear surface profile measurements, but suffers from rotation-induced phase errors in dynamic situations[7]. Comparatively, rotation-induced error in PSP is more possibly resolvable than the influence of the tooth surface discontinuities in FTP, thus PSP is commonly considered as a better or more promising selection in dynamic gear measurement[8], provided that rotation-induced errors are properly handled and processed.

In recent years, a variety of methods have been proposed to address motion-induced phase errors in dynamic 3D measurement by using PSP. These methods can be broadly categorized into four types based on the core principles:

(i) Hybrid FTP&PSP approaches apply FTP in dynamic areas to avoid motion artifacts and PSP in static areas for high accuracy. The core technical problem for this category is how to precisely segment the moving and static pixels. As a typical work, Qian's group proposed a pixel-wise motion detection strategy by using a phase frame difference method (PFDM) based on frame-to-frame FTP phase variations to distinguish dynamic and static regions[9]. Similarly, Guo et al. proposed a dual-frequency composite grating method that detects motion regions via the phase relationship between two frequencies[10]. However, the disadvantages of FTP and PSP also merge together, leading to hybrid methods not being applicable to gears.

(ii) Feature-based motion tracking methods define object movement based on physical markers or feature points, and correct phase errors by establishing pixels-dispatching model. The core technical problem is how to obtain the motion information from different objects features. Lu's team proposed a series of automated tracking algorithms such as SIFT or KCF, to perform phase compensation for single or multiple moving objects without manual intervention[11, 12]. These approaches are effective for targets with distinct features and regular motion, but perform poorly on featureless surfaces or repeated features like the gears. Wang *et al.* proposed a projection-points tracking method in phase-shifting profilometry, embedding spatially known projection points into fringe patterns and using digital image correlation to obtain displacement for phase compensation[13]. The method performs well for dynamic measurements of featureless or repetitive surfaces, but the additional markers may locally disturb fringe continuity and increase computational complexity.

(iii) The methods based on phase coding optimization aim to enhances the robustness in phase extraction, and them can be further divided into frequency- and phase-design methods. Wang et al. proposed a Hilbert transform–based method that compensates motion-induced phase errors by exploiting the doubled-frequency characteristics of object motion, effectively reducing dynamic artifacts in phase-shifting profilometry[14]. Gou's team proposed a four-step PS approach which divides PS images into two groups with a $\pi/2$ phase offset to compensate motion errors[15]. Wu et al. deduced the form of motion-induced error under linear motion assumption and proposed to suppress the error by applying equal-step algorithms[16]. These phase-coding optimization methods can effectively mitigate motion-induced errors under moderate or linear motion, but their performance degrades for nonlinear, nonuniform, or large-scale motion.

(iv) Phase compensation methods derive the mathematical model of motion-induced phase error, and improve the accuracy of the phase map calculated by normal static application. Liu et al. proposed a model-based compensation approach in which the motion-induced phase shift error is derived using the projector's pinhole model and incorporated into the phase-shifting algorithm for correction[17]. Another related method estimates the unknown motion-induced phase shifts by calculating phase differences among multiple phase maps in a measurement sequence, enabling pixel-wise correction for non-uniform or deformable motion[18]. Zhang et al. further introduced a binomial self-compensation (BSC) model that suppresses motion errors through weighted summation of successive motion-affected phase frames[19]. Although these model-based approaches improve robustness and accuracy in dynamic measurement, their effectiveness decreases under rapid motion or complex deformation.

Overall, although above methods can reduce motion-induced phase errors to some extent, they still struggle to achieve high-precision 3D reconstruction in scenarios with high dynamics, strong non-uniform motion, or complex object surfaces. Most methods implicitly assume that the magnitude of pixel-level motion between frames is small and nearly uniform, which is overly idealized and therefore inapplicable to 3D measurement of rotating gears. Recently, with the rapid development of deep learning (DL), neural network technologies have been widely applied in FPP[20-24], achieving notable progress in phase demodulation, fringe denoising, non-sinusoidal fringe analysis, and phase unwrapping. Guo et al. proposed a convolutional neural network HRU-net to directly demodulate phase from closed fringe patterns with high accuracy[25]. Reyes-Figueroa et al. proposed a modified U-Net (V-Net) to filter and normalize fringe patterns, effectively reducing noise and intensity variations[26]. Feng et al. proposed a DL-based one-to-many framework for non-sinusoidal fringe analysis, effectively compensating nonlinear and harmonic distortions[27]. Yin et al. proposed a DL-TPU network to achieve reliable and high-precision temporal phase unwrapping in FPP[28]. For reader's convenience, we also cite some most recent articles covering above directions[29-33]. Despite their remarkable performance, DL-based methods remain limited by data requirements and generalization capability.

Recently, researchers begun to explore the competence of DL in motion-induced error correction. Tan et al. developed a deep residual U-Net to achieve end-to-end correction of absolute phase [34], and Li et al. proposed a three-stream U-Net architecture that effectively suppressed motion-induced errors in unwrapped phase[35]. Wang et al. modified the U-Net by introducing residual blocks and physical priors, building a dual-network framework for single-shot super-resolved phase demodulation and unwrapping[36]. Chen et al. employed a two-stage U-Net architecture to decompose multiplexed dual-frequency fringes and unwrap phase for ultra-high-speed 3D imaging[37]. Li et al. optimized a U-Net-based architecture via NAS to create a lightweight dual-frequency network for real-time phase demodulation and 3D reconstruction[38]. Zhu et al. built a U-Net-based triple-decoder network with a physical prior to improve phase unwrapping accuracy from a single fringe pattern[39]. Through the works mentioned above, it can be seen that most deep-learning networks for motion-error correction in FPP are U-Net based, showing its strong potential for dynamic 3D measurement.

However, applying DL in dynamic FPP-gear system still faces two major challenges. Challenge (i): the intricate and repetitive tooth surface weakens the accuracy of phase recovery for current network model, especially when the angular velocity is rapidly changing. Challenge (ii): there is a lack of unified and scientific evaluation mechanisms—high-quality ground-truth 3D gear data are difficult to obtain under dynamic conditions, leading to an absence of standardized performance metrics and hindering objective comparison of results. Additionally, there are still lack high-quality ground-truth 3D gear data, which make the scientific comparison for current DL model an imprecise task. Therefore, in dynamic gear measurement, enhancing the accuracy of DL models and establishing corresponding interpretable framework for accuracy uncertainty assessment has become an urgent issue requiring resolution in this field.

This paper aims to eliminate the motion-induced phase error of FPP in dynamic gear measurement, and evaluate the uncertainty of reconstruction results simultaneously. To this end, we are pleased to introduce a novel DL model named CD-DCU-Net that is the abbreviation of Concrete Dropout–Dual Convolution U-Net. The proposed CD-DCU-Net integrates Concrete Dropout[40] for pixel-wise uncertainty estimation and employs dual 3×3 convolutions without an activation function between them in the output layer to enhance the spatial continuity of the predictions. During training process, a transfer learning[41-43] strategy is adopted: the network is firstly pretrained on simulated data and then fine-tuned on the target dataset. A weighted loss function is first-time proposed to balance phase prediction accuracy and uncertainty estimation. To facilitate efficient network training and evaluation, we construct a digital twin system based on Unity to flexibly generate diverse simulated datasets, and introduce a novel virtual-reality mixed training mode for better practice. Experimental results from rotating gear measurements demonstrate that, proposed CD-DCU-Net can achieves significant accuracy improvements over the traditional three-step PS algorithm in phase retrieval and 3D reconstruction. The corresponding output uncertainty also offers a practical and efficient evaluation metric for DL-based FPP measurement method of rotating gears.

The whole content of this paper is organized as follows: Sec. 1 presents the research background of FPP-gear measurement method and state-of-the-art works about motion-induced error correction; Sec. 2 describes the methodological principles of proposed CD-DCU-Net; Sec. 3 is the experimental validations; and Sec. 4 concludes the paper and discusses future research directions.

**2 Theory**

*2.1 Basic principle of PSP and the motion-induced gear measurement error*

In typical PSP theory, one or a series of specially designed sinusoidal fringe patterns, containing the projector pixel coordinate information and fixed phase shifts, are projected onto the object's surface. The fringe patterns will be modulated by the object's surface, thus recording the height information in captured distorted patterns by the camera[44]. By retrieving the distorted phase map from these patterns, the pixel coordinates can be extracted from camera view, i.e., the correlations between camera and projector are build. 3D reconstruction can be performed based on triangular principle[45]. The PSP conventionally assumes that the intensity reflected by the objects is linear and can be expressed as following[46]

$$I_n^c(u^c,v^c) = a^c(u^c,v^c) + b^c(u^c,v^c)\cos[\phi(u^c,v^c) + \delta_n] \qquad (1)$$

where the superscript $c$ denotes the camera, $I_n^c(u^c,v^c)$ is the gray value of the deformed phase-shifting pattern at pixel $(u^c,v^c)$, $a^c$ is the average gray value, $b^c$ is the modulation, $\phi(u^c,v^c)$ is the unwrapped phase, $\delta_n$ is the phase shift amount and $n \in \{0,1,...,N-1\}$ represents the order of phase-shifting steps, where $N$ is the total number phase-shifting steps. The wrapped phase can be calculated by the least squares solution that is typical in PSP,

$$\varphi(u^c,v^c) = \tan^{-1}\frac{M(u^c,v^c)}{D(u^c,v^c)} = \tan^{-1}\frac{-\sum_{n=1}^{N} I_n^c(u^c,v^c)\sin(\delta_n)}{\sum_{n=1}^{N} I_n^c(u^c,v^c)\cos(\delta_n)} \qquad (2)$$

where $M(u^c,v^c)$ and $D(u^c,v^c)$ represent the numerator and denominator in the arctangent function, respectively.

PSP have demonstrated its superior performance on the accuracy and robustness along a long-term period. Its pixel-independent phase-solving characteristics also offers advantages when measuring the gears with discontinuous tooth and multiple separations. However, the weakness stems from the fundamental assumption in Eq. (1), that optical signal path from 'projector→object→camera' rarely exhibits ideal linear transmission. A quantity of researches have discussed these unstable factors which may influence PSP's performance, including ambient illumination[47], projector nonlinearity[48], and camera random noise[49], etc. These factors theoretically change the gray value model of Eq. (1), causing undesired phase error the phase calculation Eq. (2).

The topic discussed in this work is a new factor which could distort the model of Eq. (1). When the object to be measured is moving or transforming, the phase-shifting method no longer performs as intended, as depicted in Fig. 1. Figure 1(a) shows one typical FPP-gear mearing system, and Fig. 1(b) illustrates the pixel correlation variation when the gear is rotating. In the geometry of imaging systems, the pixel correspondence would be different in each captured fringe pattern, i.e., one camera pixel $(x_c, y_c)$ no longer "sees" the same physical points on the gear and also the projector pixel, as the points $P_0 - x_0^p$ and $P_1 - x_1^p$ shown in Fig. 1(b). Therefore, we have to re-model the captured grayscale in Eq. (1) to include the influence caused by gear rotation. The new equation is shown as follows

$$I_n^c(u^c,v^c) = a_n^c(u^c,v^c) + b_n^c(u^c,v^c)\cos[\phi(u^c,v^c) + \delta_n] \tag{3}$$

where the average gray value $a_n^c(u^c,v^c)$, the modulation $b_n^c(u^c,v^c)$ and the phase map $\phi(u^c,v^c)$ are related to the step mark n. For convenience, we unify the phase to the same start time, and display the phase shift caused by the motion separately as $\varepsilon_n(u^c,v^c)$. The transformed Eq. (7) can be obtained as follows,

$$I_n'^c(u^c,v^c) = a_n^c(u^c,v^c) + b_n^c(u^c,v^c)\cos[\phi(u^c,v^c) + \delta_n + \varepsilon_n(u^c,v^c)]. \tag{4}$$

Therefore, the phase decoding would become an ill-posed problem, due to each captured pattern is captured with three new unknows $a_n^c(u^c,v^c)$, $b_n^c(u^c,v^c)$, and $\phi(u^c,v^c)$. If conventional PSP algorithm is directly applied to these patterns comprising the object motion information, unexpected new motion-induced phase error will appear on reconstructed gear surface, which would be harmful to the measurement accuracy. In addition, $a_n^c(u^c,v^c)$, $b_n^c(u^c,v^c)$, and $\phi(u^c,v^c)$ are related to the surface color, reflectivity, and the most important gear motion state (including spatial pose, angular velocity), making phase extraction from rotating gears appear theoretically difficulty.

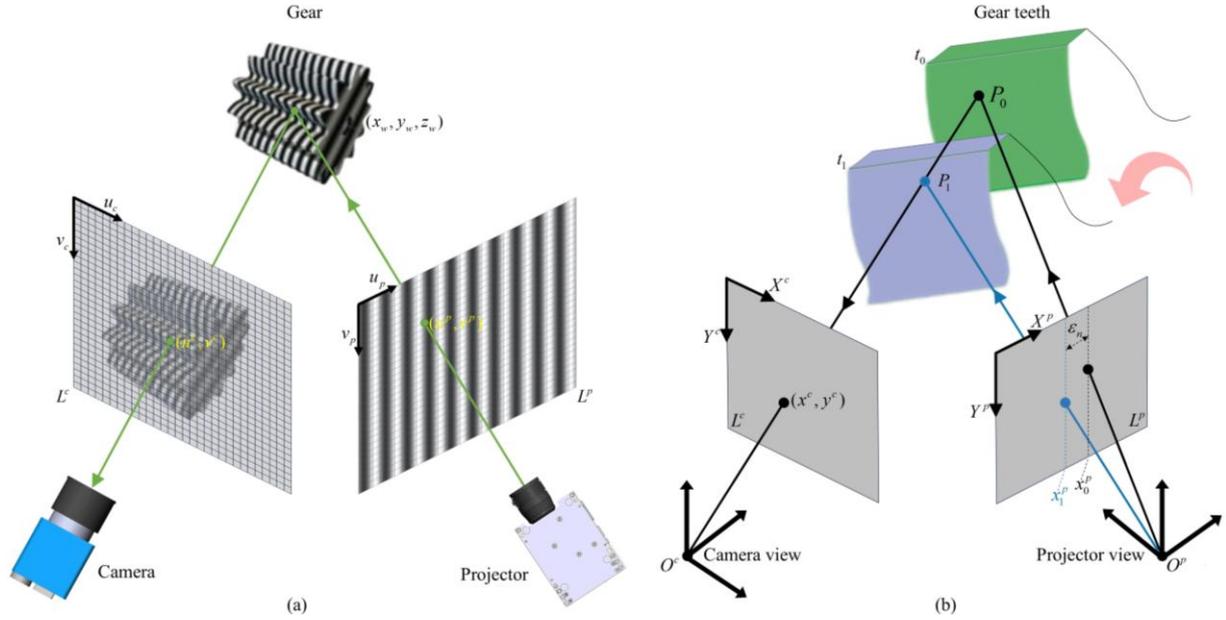

**Fig. 1.** (a) Schematic diagram of the FPP system, which consists of a camera, a projector, and the object to be measured. $L^c$ and $L^p$ represent the image plane and the projection plane, respectively. $u$ and $v$ denote the pixel coordinates on the image or projection plane, while $(x_w, y_w, z_w)$ indicate the world coordinates of a point on the gear. (b) Principle of dynamic error due to gear rotation. X, Y represent the physical coordinate axes of the projection plane or image plane. $P_0$ and $P_1$ denote the positions of a point on the gear tooth corresponding to the same pixel on the image plane at times $t_0$ and $t_1$, respectively. The unknown phase shift $\varepsilon_n$ on the projector is caused by the displacement vector from $P_0$ to $P_1$.

To reduce the number of unknown parameters and mathematically model the motion-induced phase error, we adopt three assumptions. (i) The background intensity is uniform, thus $a_n^c(u^c,v^c)$ can be represented by one parameter, i.e., $a_n^c(u^c,v^c) = a$. (ii) A small local region on object surface has the same reflectivity, thus $b_n^c(u^c,v^c)$ can be represent by one parameter, i.e., $b_n^c(u^c,v^c) = b$. (iii) The phase shift is assumed to be sufficiently small for the small-angle approximations $\sin\varepsilon_n \approx \varepsilon_n$ and $\cos\varepsilon_n \approx 1$ to hold, and the motion corresponding to each pixel is uniform and can be simplified to a linear change, i.e., $\varepsilon_n = (n-1)\varepsilon$. Based on the first two assumptions, we know that the background term $a$ and the reflectivity term $b$ are constant between consecutive phase-shifted frames. This greatly simplifies the modeling of the phase error. From Eq. (2) the ideal phase can be expressed, and the motion-induced phase error is obtained by replacing $I_n^c(u^c,v^c)$ with the ideal intensities $I_n'^c(u^c,v^c)$ in Eq. (4). Thus, the phase error $\Delta\varphi(u^c, v^c)$ can be expressed as the erroneous phase $\varphi_{ER}(u^c, v^c)$ minus the ideal phase $\varphi_{GT}(u^c, v^c)$, and the result is

$$\Delta\varphi(u^c, v^c) \tan^{-1} \frac{-\sum_{n=1}^{N} I_n'^c(u^c, v^c)\sin(\delta_n)}{\sum_{n=1}^{N} I_n'^c(u^c, v^c)\cos(\delta_n)} - \tan^{-1} \frac{-\sum_{n=1}^{N} I_n^c(u^c, v^c)\sin(\delta_n)}{\sum_{n=1}^{N} I_n^c(u^c, v^c)\cos(\delta_n)}. \quad (5)$$

Eq. (5) is rather complicated, and directly expanding its trigonometric terms is prone to computational errors. To simplify the derivation, we convert Eq. (5) into the complex domain using Euler's identity (see Appendix). The resulting compact form is

$$\Delta\varphi(u^c, v^c) = \arctan \frac{\sum_{n=1}^{N}\sin\varepsilon_n - \cos(2\varphi_{GT})\sum_{n=1}^{N}\sin(2\delta_n + \varepsilon_n) - \sin(2\varphi_{GT})\sum_{n=1}^{N}\cos(2\delta_n + \varepsilon_n)}{\sum_{n=1}^{N}\cos\varepsilon_n + \cos(2\varphi_{GT})\sum_{n=1}^{N}\cos(2\delta_n + \varepsilon_n) - \sin(2\varphi_{GT})\sum_{n=1}^{N}\sin(2\delta_n + \varepsilon_n)}. \quad (6)$$

For brevity, the arctangent expression is presented without explicit pixel-wise notation; however, all variables remain functions of the pixel coordinates. Based on the first part of assumption (iii), Eq. (6) can be simplified to

$$\Delta\varphi(u^c, v^c) \approx \frac{\sum_{n=1}^{N}(1 - \cos(2\varphi_{GT} + 2\delta_n))\varepsilon_n}{N}. \quad (7)$$

By further incorporating the latter part of assumption (iii), where the inter-frame phase shift is represented linearly, Eq. (7) can be further simplified to

$$\Delta\varphi(u^c, v^c) \approx \frac{\varepsilon}{N}\left[\frac{N(N-1)}{2} - \sum_{n=1}^{N}(n-1)\cos(2\varphi_{GT} + 2\delta_n)\right]. \quad (8)$$

Under the three assumptions, the motion-induced phase error in Eq. (7) and (8) can be uniformly expressed as

$$\Delta\varphi(u, v) = A(u, v)\cos(2\varphi_{GT}(u, v) + \Phi) + B, \quad (9)$$

where $A(u, v)$, $\Phi$, and $B$ are coefficients determined by $\varepsilon_n$ and $\delta_n$. This general form indicates that the phase error essentially corresponds to a second-harmonic cosine modulation of the ground-truth phase $\varphi_{GT}$, without introducing any additional spatial-frequency components.

However, the three assumptions above are overly idealized and clearly do not hold in dynamic gear measurement. First, whether the gear undergoes uniform or non-uniform motion, the resulting unknown phase shift cannot remain constant; in fact, the large height variations on the tooth surface make the phase-shift behavior even more complex and difficult to predict. Second, during gear rotation, self-occlusion inevitably occurs between the tooth tips and roots, as illustrated in ROIs 1 and 2 of Fig. 2. Such significant height differences not only cause severe distortion of the fringe patterns, but also lead to pronounced discrepancies in the phase shifts between occluded and non-occluded regions, resulting in spatially inconsistent and sometimes substantially different phase estimates. To address these challenges and achieve robust dynamic gear measurement, we introduce a deep-learning-based method to compensate for motion-induced phase errors and reduce estimation uncertainty, as detailed in the next section.

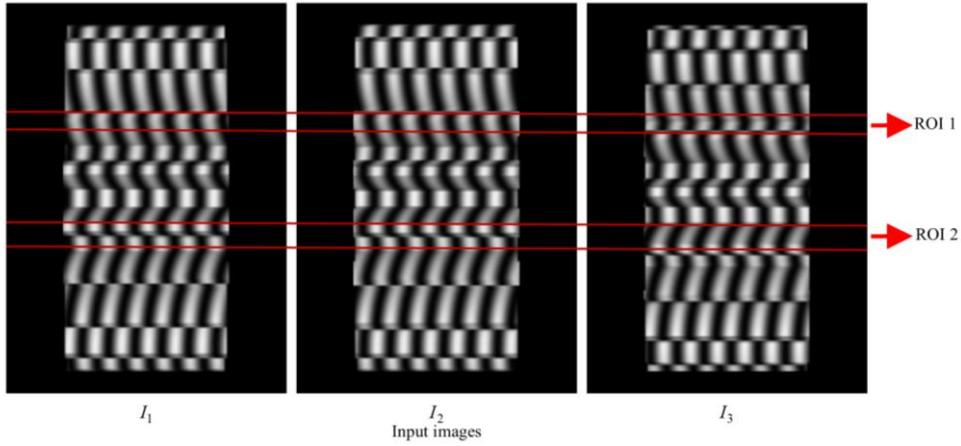

**Fig. 2.** shows a schematic of the occlusion effect. During dynamic gear measurement, as the gear rotates, certain regions (such as ROI 1 and ROI 2) are occluded or revealed in different frames, resulting in discontinuities in the measurement results.

*2.2 CD-DCU-Net architecture*

The CD-DCU-Net architecture employs a symmetric encoder-decoder structure to predict wrapped phase and quantify uncertainty in dynamic gear three-step phase-shifted fringe patterns. The network processes three consecutive phase-shifted grayscale images (800×800×3) and outputs both phase predictions and corresponding uncertainty estimates.

In the encoder path, the basic feature extraction unit (Double Conv) consists of two consecutive 3×3 convolutions, each followed by batch normalization and ReLU activation. Feature channels expand progressively (64→128→256→512→1024) through 2×2 max pooling operations (stride = 2). To quantify model uncertainty, we strategically integrate Concrete Dropout modules at key positions in the network. Specifically, Concrete Dropout layers are applied after each down sampling operation in the encoder path and after each up sampling in the decoder path, with feature channels of 64, 128, 256, 512, 1024 and 512, 256, 128, 64, respectively. Each Concrete Dropout layer includes a dedicated 3×3 convolution to process dropped-out features, allowing dropout rates to be adaptively learned during training rather than fixed as hyperparameters. The decoder path employs 2×2 transposed convolutions (stride = 2) for up sampling, with classic U-Net skip connections preserving spatial information between corresponding layers.

A key innovation is introduced in the output layer (Out-Conv), where two consecutive 3×3 convolutions are employed. Unlike the Double Conv blocks in the main body of the network, which utilize batch normalization and ReLU activation to enhance feature representation and training stability, the output layer omits any normalization or activation functions after the convolutions. The first 3×3 convolution maintains the same number of channels as the preceding layer, and the second 3×3 convolution then reduces the channels to the required output channels, with no activation function applied. This design helps stabilize the training process and, more importantly, preserves the continuity and numerical precision of the final output, which is crucial for accurate regression of continuous-valued quantities. Two consecutive 3×3 convolutions are equivalent to a single 5×5 convolution in terms of receptive field, and allow for flexible control of the intermediate channel number, resulting in more adaptable parameter usage and feature processing.. This structure is particularly effective for continuous regression tasks, as the expanded receptive field enables the network to consider broader spatial context simultaneously, resulting in smoother and more continuous predictions. This is especially important for wrapped phase, which is inherently a continuously varying physical quantity, as the network can more accurately capture subtle relational changes between adjacent pixels, reducing discontinuities and local mutations in predictions. The final layer outputs dual-channel means (M and D) and their respective variances ($\sigma_M^2$ and $\sigma_D^2$), effectively addressing the wrapped phase prediction task while quantifying prediction reliability.

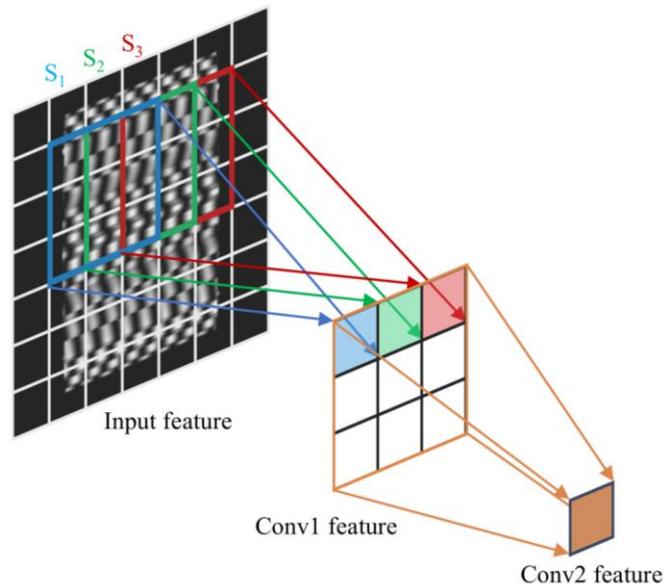

**Fig. 3.** Schematic diagram of the equivalent receptive fields of two consecutive 3×3 convolutional layers and a single 5×5 convolutional layer. From left to right, the figure illustrates the input feature map, the intermediate feature map after the first 3×3 convolution (conv1 feature), the output feature map after the second 3×3 convolution (conv2 feature), and the output feature map of a single 5×5 convolutional layer. $S_1$, $S_2$, and $S_3$ in the figure represent the sequential sliding positions of the convolution kernels on the feature map. By comparing the receptive field coverage of the two structures on the feature map, the diagram visually demonstrates the equivalence in spatial feature extraction between two stacked 3×3 convolutions and a single 5×5 convolution.

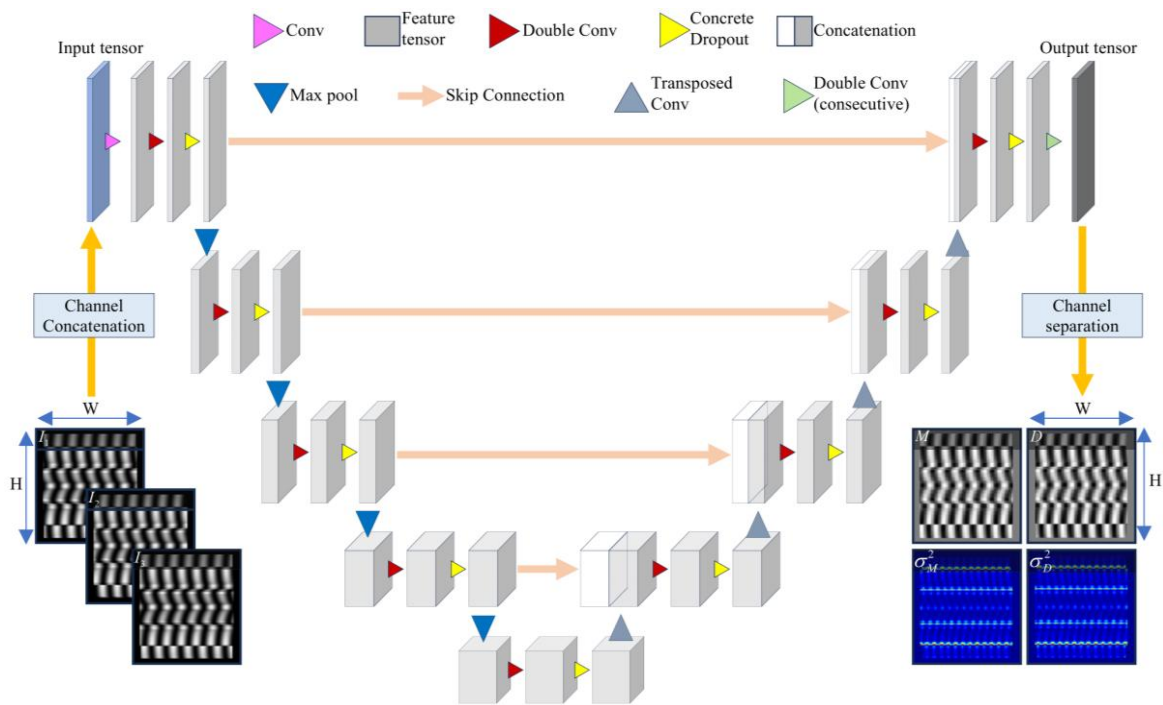

**Fig. 4.** CD-DCU-Net architecture. The model adopts a symmetric encoder-decoder structure, with three consecutive phase-shifted grayscale images (800×800×3) as input. The encoder achieves multi-scale feature extraction through four down-sampling operations and feature channel expansion (3→64→128→256→512→1024); the decoder uses transposed convolution for up-sampling, fusing multi-scale information through skip connections. The model innovatively introduces the Concrete Dropout mechanism for uncertainty modeling, and adopts a dual 3×3 convolution design in the output layer, finally outputting phase predictions and their uncertainty estimates.

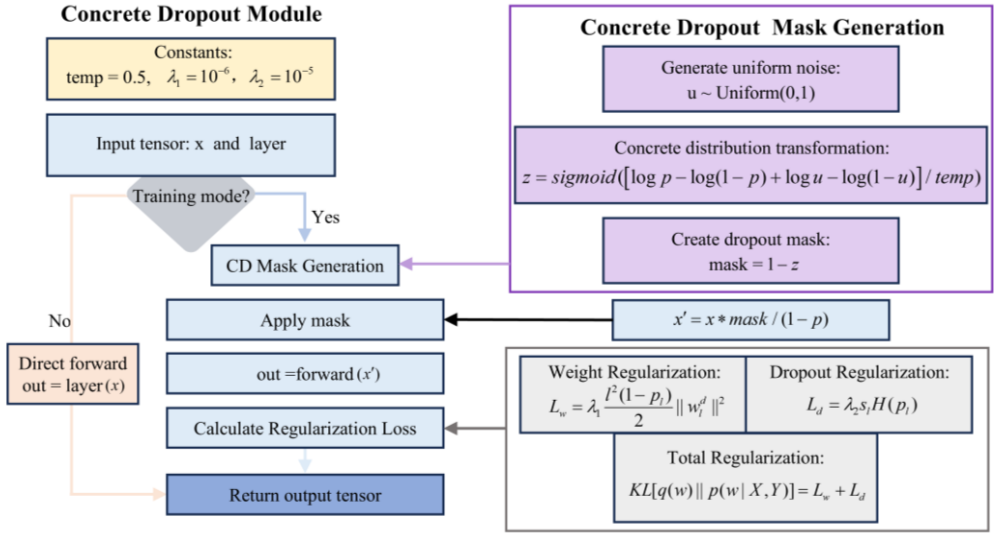

**Fig. 5.** Workflow diagram of the Concrete Dropout module with learnable dropout rates.

*2.3 Probability distribution modeling*

To model prediction uncertainty, we implement a Bayesian Neural Network (BNN) using Concrete Dropout[30, 50]. This approach approximates Bayesian inference by treating network weights as probability distributions. Our system processes fringe image sets $X = \{x^k\}_{k=1}^{K}$, where each $x^k = \{x^{k1}, x^{k2}, x^{k3}\}$ consists of a 3-step PS group. The network is trained on ground truth labels $Y = \{y^k\}_{k=1}^{K}$, where each $y^k$ contains ideal numerator and denominator values. The network weights $w$ follow a probabilistic distribution, allowing us to capture both data uncertainty (noise inherent in input data) and model uncertainty (weight confidence learned from training data). For an input $x$, the predictive distribution integrates these two types of uncertainty as shown in the following equation:

$$p(y|x, X, Y) = \int \overbrace{p(y|x, w)}^{Data\ uncertainty} \cdot \underbrace{p(w|X, Y)}_{Model\ uncertainty}\ dw \qquad (8)$$

To model data uncertainty in pixel-level predictions, we assume $N$ independent pixels in the output $y^k$. The likelihood factors as[30]:

$$p(y^k | x^k, w) = \prod_{i=1}^{N} p(y_i^k | x^k, w) \qquad (9)$$

Each pixel follows a temperature-scaled Gaussian distribution:

$$p(y_i^k | x^k, w) = \frac{1}{\sqrt{2\pi}\sigma_i^k} \exp\left[-\frac{(y_i^k - \mu_i^k)^2}{2(\sigma_i^k)^{2\alpha}}\right] \qquad (10)$$

where $\mu_i^k$ is the predicted mean of pixel, $\sigma_i^k$ is the standard deviation of pixel, and temperature constant $\alpha$ controls distribution width ($\alpha > 1$: wider/more uncertain, $\alpha > 1$: narrower/sharper).

The model learns by minimizing the negative log-likelihood:

$$-\frac{1}{N}\log p(y^k \mid x^k, w) = \frac{1}{N}\sum_{i=1}^{N}\left[\frac{1}{2(\sigma^k)^2}\mid y^k - \hat{y}^k \mid^2 + \frac{1}{2}\log(\sigma^k)^{2\alpha}\right] \quad (11)$$

Here $y$ is the ground truth, $\hat{y}$ is the BNN's prediction, and $\sigma^2$ is variance. Crucially, this allows simultaneous unsupervised learning of both predictions and uncertainty estimates—no true variance labels required. We set $\alpha = 10^{-4}$ in this work, optimally balancing the loss contributions between the mean prediction and variance estimation.

To model model uncertainty[30, 51], we employ a Bayesian neural network with Concrete Dropout. In this framework, the dropout rate of each layer is automatically learned during training, enabling more accurate uncertainty estimation. Variational inference forms the core of this method, where we approximate the intractable true posterior distribution $p(w \mid X, Y)$, with a variational distribution $q(w)$. The predictive distribution for a new input $x$ is approximated through Monte Carlo integration:

$$p(y \mid x, X, Y) \approx \int p(y \mid x, w)q(w)dw \approx \frac{1}{T}\sum_{t=1}^{T} p(y \mid x, w^{(t)}) \quad (12)$$

Here, we sample weights $w^{(t)}$ from $q(w)$ and average predictions to approximate the output distribution for $x$.

The network is trained by minimizing the following loss function:

$$L_{BNN} = -\frac{1}{N}\sum_{i=1}^{N}\log p(y^k \mid x^k, w) + KL[q(w) \parallel p(w \mid X, Y)] \quad (13)$$

The variational distribution factorizes as a product across layers[40]:

$$q(w) = \prod_{l=1}^{L} q_{w_l^d}(w_l) \quad (14)$$

$$q_{w_l^d}(w_l) = w_l^d \cdot diag\left[Bernoulli(1 - p_l)^{s_l}\right] \quad (15)$$

Here, $L$ is the number of dropout layers, $w_l^d$ denotes the deterministic weight of the $l$-th layer, $w_l$ is the random weight after applying dropout, $p_l$ is the dropout rate, and $s_l$ is the dimensionality of $w_l$

The KL divergence term approximates to[30]:

$$KL[q(w) \parallel p(w \mid X, Y)] = \sum_{l=1}^{L} KL\left[\lambda_1 \frac{p_l(1-p_l)}{2} \parallel w_l^d \parallel^2 - \lambda_2 s_l H(p_l)\right] \quad (16)$$

$$H(p_l) = -p_l \log p_l - (1 - p_l)\log(1 - p_l) \quad (17)$$

where $\lambda_1$ and $\lambda_2$ are regularization parameters for the weights and dropout rates, respectively, $H(p_l)$, is the entropy of the Bernoulli random variable.

The key innovation of Concrete Dropout is that it replaces the discrete Bernoulli variables in traditional dropout with a continuous Concrete distribution, allowing the dropout mask to take values in the interval (0,1). The expression is as follows[30]:

$$z_l = sigmoid\left(\frac{1}{\beta}[\log p_l - \log(1 - p_l) + \log u - \log(1 - u)]\right) \quad (18)$$

Here, $z_l$ denotes the continuous dropout mask for the $l$-th layer, which weights the outputs of neurons in this layer; β=0.5 is the temperature parameter that regulates the relaxation degree of the distribution; u~Uniform(0,1) is a random variable following the uniform distribution. This continuous design enables the dropout probability to be optimized via gradient descent, providing a more flexible parameter adjustment mechanism for model uncertainty quantification, and the Concrete Dropout module's specific implementation process is shown in Figure 5

*2.4 Uncertainty Quantification of Model Predictions via Monte Carlo Sampling*

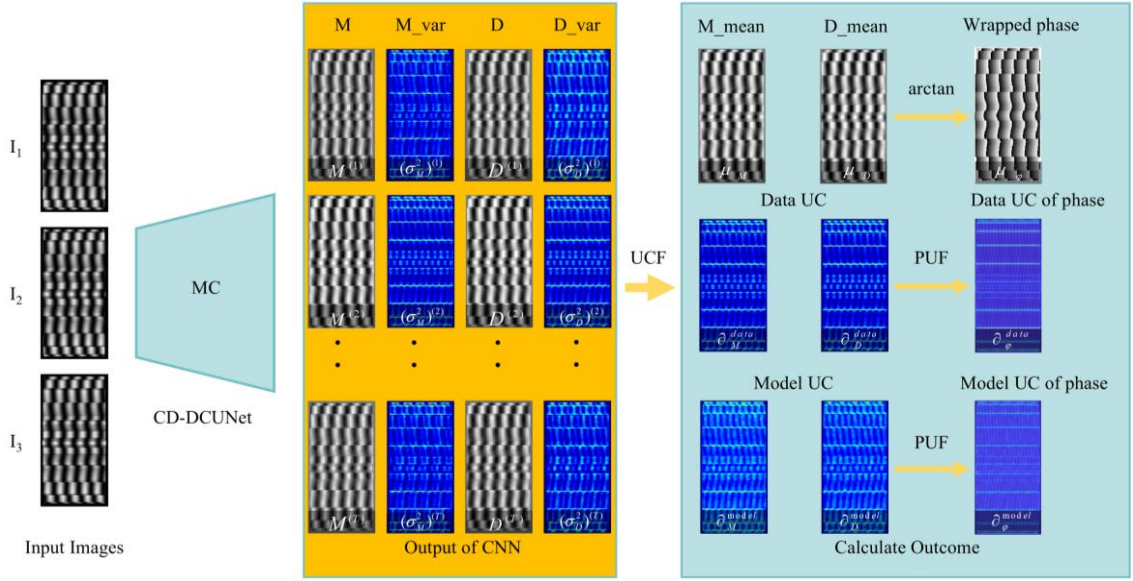

**Fig. 6**. Flowchart of uncertainty quantification via Monte Carlo (MC) sampling. The method first performs T forward predictions on the input fringe pattern image. Each prediction yields a set of output parameters via the concrete dropout mechanism: the sine term $M^{(t)}$ and its variance $(\sigma_M^2)^{(t)}$, the cosine term $D^{(t)}$ and its variance $(\sigma_D^2)^{(t)}$. Substituting the computed sine mean $\mu_M$ and cosine mean $\mu_D$ into the arctangent function gives the wrapped phase $\mu_\varphi$. Finally, through uncertainty calculation formulas (UCF), the model uncertainty (UC) $\partial_M^{model}$, $\partial_D^{model}$ and data uncertainty $\partial_M^{data}$, $\partial_D^{data}$ of the sine and cosine terms are obtained, and through propagation of uncertainty (PUF) formulas, the phase uncertainties (UC) $\partial_\varphi^{model}$, $\partial_\varphi^{data}$ and are further derived.

The uncertainty estimation in this work is based on Monte Carlo (MC) sampling. The model and data uncertainties are obtained according to the uncertainty calculation formulas (UCF)[30, 51], and the corresponding phase uncertainties are further derived through the propagation of uncertainty (PUF) formulas[30]. Finally, we propose the following formulation to combine model and data uncertainties into the total phase uncertainty:

$$\hat{\partial}_\varphi^{total} = \sqrt{\left(\hat{\partial}_\varphi^{model}\right)^2 + \left(\hat{\partial}_\varphi^{data}\right)^2} \qquad (19)$$

Where $\partial_\varphi^{model}$ is the model uncertainty of the phase, $\partial_\varphi^{model}$ is the data uncertainty of the phase.

*2.5 Construction of Virtual Fringe Projection System*

To accurately construct a virtual fringe projection system in Unity based on the real system, it is necessary to clarify the correspondence between the intrinsic and extrinsic parameters obtained from calibration and the control parameters used in Unity.

The intrinsic parameters of the real system, obtained after calibration, include the focal length $f$, sensor size $w \times h$, and principal point $(u_0, v_0)$. These parameters describe the internal geometric and optical characteristics of the camera or projector.

The extrinsic parameters, represented by the rotation matrix $(R)$ and the translation vector $(T)$, define the position and orientation of the camera (or projector) in the world coordinate system.

The mapping and conversion relationships between real system parameters and Unity input parameters are summarized in Table1. This table lists the correspondence of focal length, sensor size, principal point, field of view, resolution, and extrinsic pose, as well as how they are implemented in Unity. Notably, in real systems, the field of view (FOV) is not a direct physical parameter but is calculated from the focal length and sensor size. In Unity, however, the FOV must be explicitly set for the Projector component.

The position of the camera (or projector) center in the world coordinate system is calculated as：

$$C = (X_c, Y_c, Z_c)^T = -R^{-1}T \tag{20}$$

where $C$ represents the camera (or projector) center in world coordinates, $R$ is the rotation matrix, and $T$ is the translation vector obtained from calibration.

The transpose of the rotation matrix $R$ can be decomposed into Euler angles compatible with Unity's configuration according to:

$$R^T = R_{z(\phi)}R_{x(\theta)}R_{y(\psi)} = \begin{bmatrix} \cos\phi & -\sin\phi & 0 \\ \sin\phi & \cos\phi & 0 \\ 0 & 0 & 1 \end{bmatrix} \begin{bmatrix} 1 & 0 & 0 \\ 0 & \cos\theta & -\sin\theta \\ 0 & \sin\theta & \cos\theta \end{bmatrix} \begin{bmatrix} \cos\psi & 0 & \sin\psi \\ 0 & 1 & 0 \\ -\sin\psi & 0 & \cos\psi \end{bmatrix} \tag{21}$$

where $z(\phi)$, $x(\theta)$, $y(\psi)$ are the rotations around the z, x, y axes, respectively[52].

**Table. 1** summarizes the mapping and conversion relationships between real system parameters and Unity input parameters:

| Real System Parameter | Unity Control Parameter | Mapping/Conversion | Description |
|---|---|---|---|
| Focal length ($f$) | Focal length (Camera, mm) | Direct input | For Camera: can be set in Physical Camera mode |
| Sensor size ($w, h$) | Sensor size (Camera, mm) | Direct input | For Camera: can be set in Physical Camera mode |
| Principal point ($u, v$) | Lens shift (Camera) | Can be approximated using lens shift | Offset from image center |
| Resolution (pixels) | Resolution (Camera/Projector) | Set to match real system | Image output size in pixels |
| Focal length ($f$), sensor width ($w$) | Field of View (FOV, Projector, deg) | $FOV = 2\arctan\left(\dfrac{w}{2f}\right)$ | For Projector: FOV is calculated for Unity only |
| Rotation matrix ($R$), translation ($T$) | Transform (Position, Rotation) | Use calibration results; see formulas above | Defines pose in world coordinates |

**3.validation**

*3.1 System Construction*

Based on the dataset generation method described above, we built both a physical and a virtual FPP system in Unity for data collection. First, we calibrated the physical system using Zhang's calibration method to obtain the intrinsic and extrinsic parameters, which were then transferred to Unity. The extrinsic parameters determine the positions of the projector and camera in the world coordinate system and can be calculated using Equations (6) and (7). The intrinsic parameters define the internal optical characteristics of the camera and projector. In Unity, the camera's focal length can be directly set, while the projector lacks a direct focal length setting but has FOV and PSS (sensor size). The Field of View, measured in degrees, refers to the angular extent of the projector's visual field, defining the coverage from the projection origin. PSS represents the physical dimensions of the projector's virtual sensor in millimeters, analogous to the photosensitive element size in cameras and defining the actual size of the projection plane. The projector's focal length can be derived from Equation (38).The collaboration between the projector and point light source is crucial for structured light projection. The point light source acts as the emission origin, while the projector controls the spatial distribution of light. Setting the shadow softening parameter of the point light source to 0 simulates an ideal point source, producing sharp projection edges. This configuration ensures the projected pattern remains focused at any distance without manual focal length adjustment. Using the dataset generation method described above, data collection can be performed in both virtual and physical systems.

$$f_p = (ProSensorsSize / 2) / \tan((FOV / 2)\pi / 180) \tag{22}$$

where $f_p$ denotes the equivalent focal length of the projector (unit: millimeter), PSS represents the physical size of the projector's virtual sensor (unit: millimeter), and FOV indicates the field of view of the projector (unit: degree).

*3.2 Dataset Generation strategy*

To simulate the single rotational characteristics of gears, this paper uses linear interpolation to model both uniform and non-uniform motion processes. For each gear model with a given pitch angle (30°, 18°, or 12°), multiple initial starting points were set at 2° intervals within one pitch angle range (resulting in 15, 9, and 6 starting points, respectively). At each starting point, the gear was rotated with a fixed step size of 0.1°, producing 18 equally spaced angular positions. At each of these positions, 18 phase-shifting fringe images were captured, resulting in a total of 18 × 18 = 324 images per starting point. For dynamic error modeling, each three-step phase-shifting input was constructed by selecting one image from each of the three columns corresponding to phase shifts of 0, 2π/3, and 4π/3 (as indicated by the red boxes in Fig. 6). These three images had to be selected from three different angular positions in strictly ascending order, so as to reflect the continuous motion of the gear and accurately model dynamic error effects.

Four gear models were used in this study: three simulated models with pitch angles of 30°, 18°, and 12°, and one actual gear with a pitch angle of 18°. The simulated gears were assigned 15, 9, and 6 initial positions, respectively, based on uniform 2° intervals. The actual gear was measured at 4 different initial rotation angles, with 816 image pairs acquired per angle. Altogether, the combined dataset contained (15 + 9 + 6 + 4) × 816 = 27,744 samples. These data were split into training and validation sets in an 80%/20% ratio. The validation set included representative samples of both uniform and non-uniform motion patterns to support comprehensive performance evaluation.

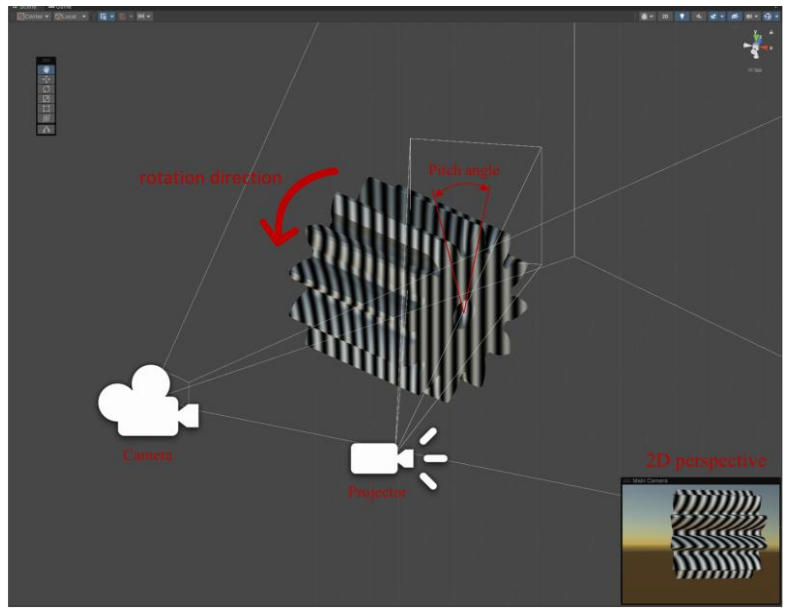

**Fig. 7.** Schematic diagram of the virtual data acquisition system in Unity, illustrating the spatial layout and relationships among the camera, projector, and rotating gear in the simulation environment. The red annotations indicate key elements including the rotation direction, pitch

angle, camera, projector, and 2D perspective view of the gear. This virtual setup enables dynamic acquisition of fringe pattern images during gear rotation, facilitating dataset generation for subsequent model training and evaluation.

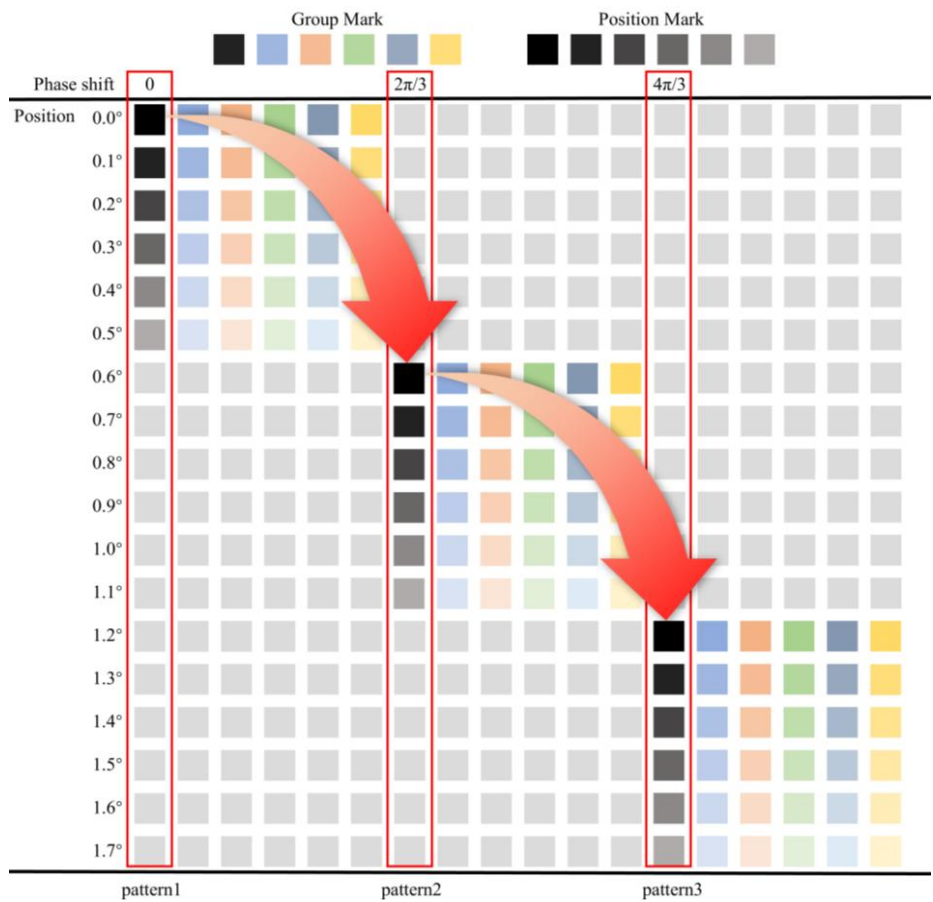

**Fig. 8.** Schematic diagram of the selection of three-step phase-shifting inputs for dynamic error modeling. Each column corresponds to a different phase shift. For each input, one image is selected from each of the three columns at three different angular positions (strictly in ascending order, as indicated by the red boxes), reflecting the continuous motion of the gear. 'Group Mark' indicates the phase shift grouping, and 'Position Mark' denotes different angular positions. The black squares illustrate one typical selection method shown in the figure. In practice, each three-step phase-shifting input can be selected from all eligible combinations, with a total of 816 possible combinations; only one is shown here.

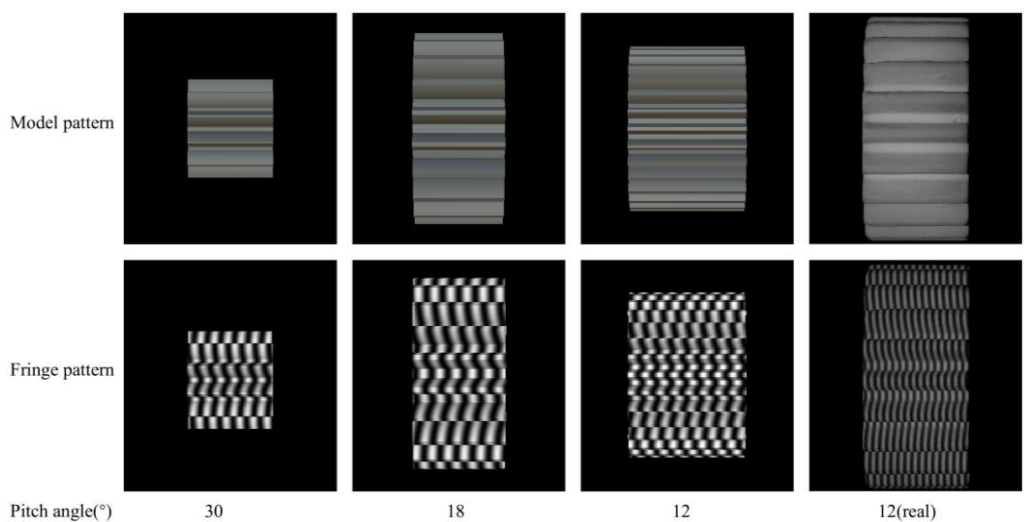

**Fig. 9.** Sample images from the gear fringe projection dataset, showing model patterns and corresponding fringe patterns for gears with different pitch angles and sizes. The dataset consists of both simulated and real gear images. The original captured images have a resolution of 2048×2448

pixels and were cropped to 1360×1232 pixels for training purposes. The three groups on the left show simulation data for gears with pitch angles of 30°, 18°, and 12°, while the group on the right presents real captured images of a gear with a pitch angle of 18°.

*3.3 Network Training*

Due to the high cost, long acquisition time, and limited availability of real dynamic gear measurement data, while simulated data can be rapidly and abundantly generated, we adopt a unified training strategy based on pretraining on simulated data. Specifically, model parameters obtained from pretraining on simulated data are used to initialize all experimental networks. This pretrained model only needs to be trained once and can be reused for all subsequent experiments without retraining for each new dataset. Unless otherwise specified, all networks in this work follow this training strategy. This approach is widely adopted in similar tasks due to its efficiency and ease of reuse. The training process consists of two stages: first, a deterministic model without Dropout is trained to obtain an initial parameter distribution; then, Concrete Dropout is introduced for formal training. Concrete Dropout layers are configured with a weight regularizer of $10^{-5}$ and a dropout regularizer of $10^{-6}$, with initial dropout rates ranging from 0.1 to 0.4. To prevent overfitting while supporting mean and variance prediction, the loss function adopts a negative log-likelihood combined with Concrete Dropout regularization. The Adam optimizer is used with an initial learning rate of 0.001, decaying to 90% every 20 epochs. The batch size is set to 16, and training runs for up to 100 epochs with early stopping if the validation loss does not improve for 10 consecutive epochs. The input image size is 2048×2448, and it undergoes four down sampling operations, reducing the resolution sequentially to 1/2, 1/4, 1/8, and 1/16 of the original size. The decoder path restores the resolution through four up sampling stages. The network outputs four channels: the mean of *M*, the mean of *D*, the variance of *M*, and the variance of *D*. The variance channels are passed through a soft plus function with an added constant of 0.001 to ensure positive values and avoid numerical instability. The training flowchart of the network is shown below.

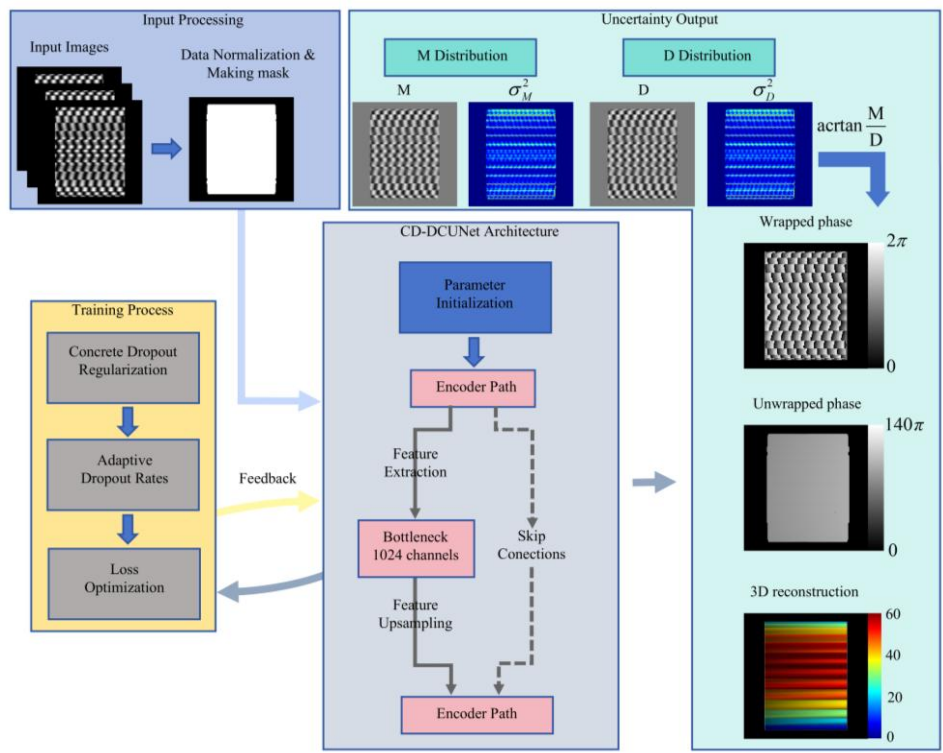

**Fig. 10.** Overall framework of the proposed method. The framework mainly consists of four functional modules: (1) Input Preprocessing Module that normalizes three-step phase-shifting images and generates masks; (2) U-Net-based DL Architecture achieving multi-scale feature extraction and fusion through an encoder-decoder structure with skip connections, where the bottleneck layer is set to 1024 channels for enhanced feature representation; (3) Training Optimization Module introducing the Concrete Dropout regularization mechanism to improve

model robustness via adaptive dropout rate adjustment and loss function optimization; (4) Uncertainty Output Module predicting the means and variances of sine term (M) and cosine term (D), obtaining wrapped phase through arctangent operation on M and D, and finally completing phase unwrapping and 3D reconstruction.

*3.4 Verification of Network Prediction Accuracy*

Before conducting comparative studies with other networks, we first designed an experiment to verify the effectiveness of the proposed training strategy. Specifically, the proposed network was trained with two different initialization schemes: Kaiming initialization and weights pretrained on simulated data. The results show that the accuracy change on simulated data is minor, while a significant improvement, exceeding 50%, is achieved on real noisy data. This indicates that simulation-based pretraining can substantially enhance the robustness and generalization ability of the model under practical conditions, thereby validating the rationality of the proposed training strategy. The experimental results are shown in Figure 11.

To comprehensively evaluate the effectiveness of the proposed method, this paper selects three typical network architectures for comparative experiments. First, as a classic encoder–decoder framework, the traditional U-Net[53] uses 1×1 convolution in the output layer to perform feature mapping operations. Although its structure is simple, its performance is limited when handling tasks with strong spatial correlation such as stripe projection. Second, the TS-U-Net[35] (temporal three-stream neural network) inputs stripe image sequences into multiple parallel branches in a grouped manner and introduces 3×3 convolution in the output layer to enhance local feature extraction. However, its receptive field and inter-branch information interaction remain insufficient. Finally, the nonlinear mapping method based on the DR-U-Net[34] (deep residual U-Net), although it improves feature representation capability via residual-block stacking, is prone to feature redundancy and incurs excessive computational overhead. To ensure the fairness of the experiments, all models are trained on the same dataset with identical hyperparameter settings (random seeds, learning rates, batch sizes). Meanwhile, an early-stopping mechanism is adopted (training terminates if the validation-set loss does not improve for 30 consecutive epochs) to ensure each model reaches its optimal performance state. The prediction accuracies of the wrapped phases under uniform and non-uniform gear motions are compared on the validation set, with the angular increments for uniform motion set to 0.6° between frames 1–2 and 2–3, and for non-uniform motion to 0.1° and 1.6° respectively. Figure 12 shows the training and validation loss comparison for each method, while Figures 13 and 14 present the wrapped phase prediction results under uniform and non-uniform rotation, respectively. It can be observed that the proposed CD-DCU-Net achieves at least a 5%–30% improvement in prediction accuracy over the other three networks.

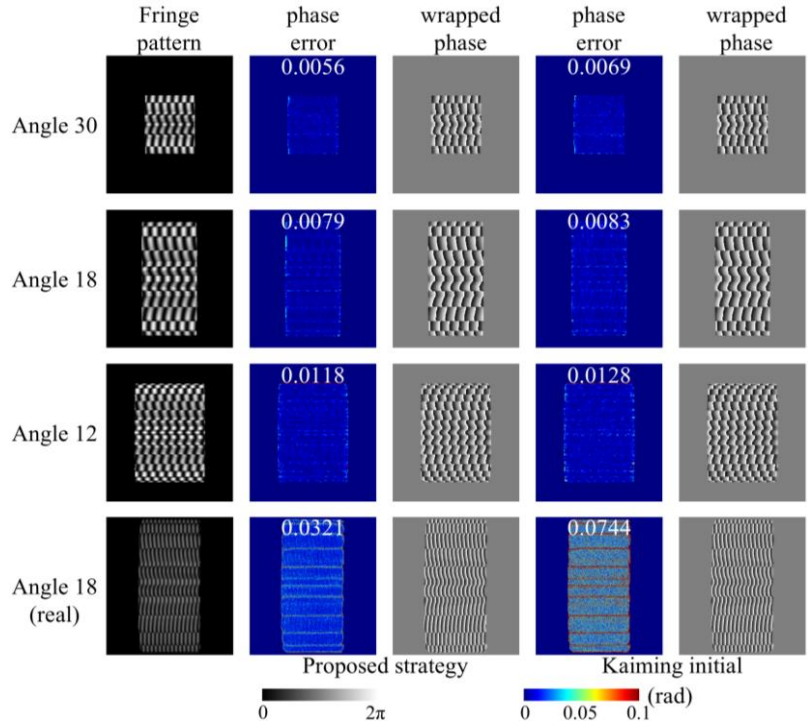

**Fig. 11.** Comparison of wrapped phase prediction for four gear samples (pitch angles of 30°, 18°, 12°, and real gear with 18°) under rotation, evaluated on a randomly selected validation set. The first column shows the fringe patterns. The remaining four columns are grouped in pairs: the first pair (proposed strategy) shows the wrapped phase RMSE and predicted wrapped phase for the model initialized with simulation-pretrained weights; the second pair (Kaiming initialization) shows the corresponding results for the model initialized with Kaiming initialization.

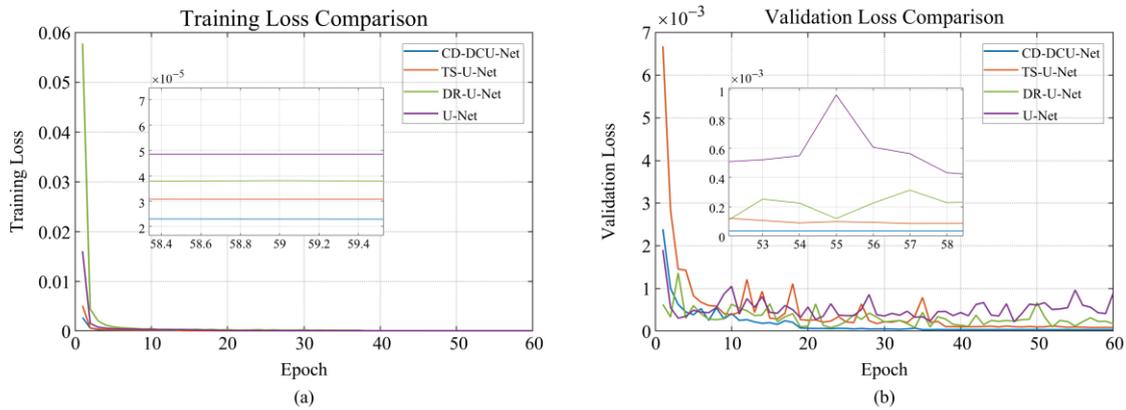

**Fig. 12.** Comparison of MSE Loss Values during Training Processes of Different Methods (a) Training set loss curves and (b) validation set loss curves. It can be observed from the figures that the proposed method (blue line) converges rapidly in the early training stage and finally reaches the lowest loss value. In contrast, the three-stream neural network (red line), residual U-Net (purple line), and traditional U-Net (green line) show slower convergence speeds and higher final loss values. The results on the validation set also indicate that the proposed method has better generalization performance, with its loss curve being more stable and having smaller fluctuations.

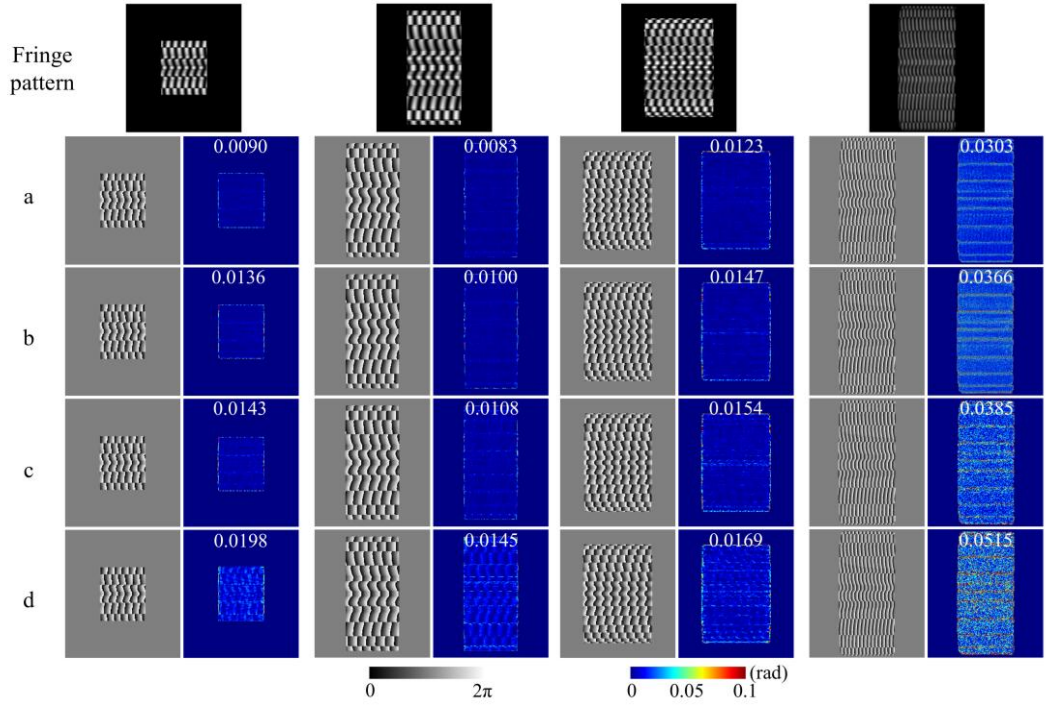

**Fig. 13.** Comparison of wrapped-phase prediction results by four networks under uniform motion. Rows a–d corresponds to the four methods (a: CD-DCU-Net, b: TS-U-Net, c: DR-U-Net, d: U-Net), and the topmost row shows the original fringe patterns of different gears. For each gear, the two columns below show the predicted wrapped phase (left) and the error heatmap with respect to the ground truth (right), with RMSE values annotated.

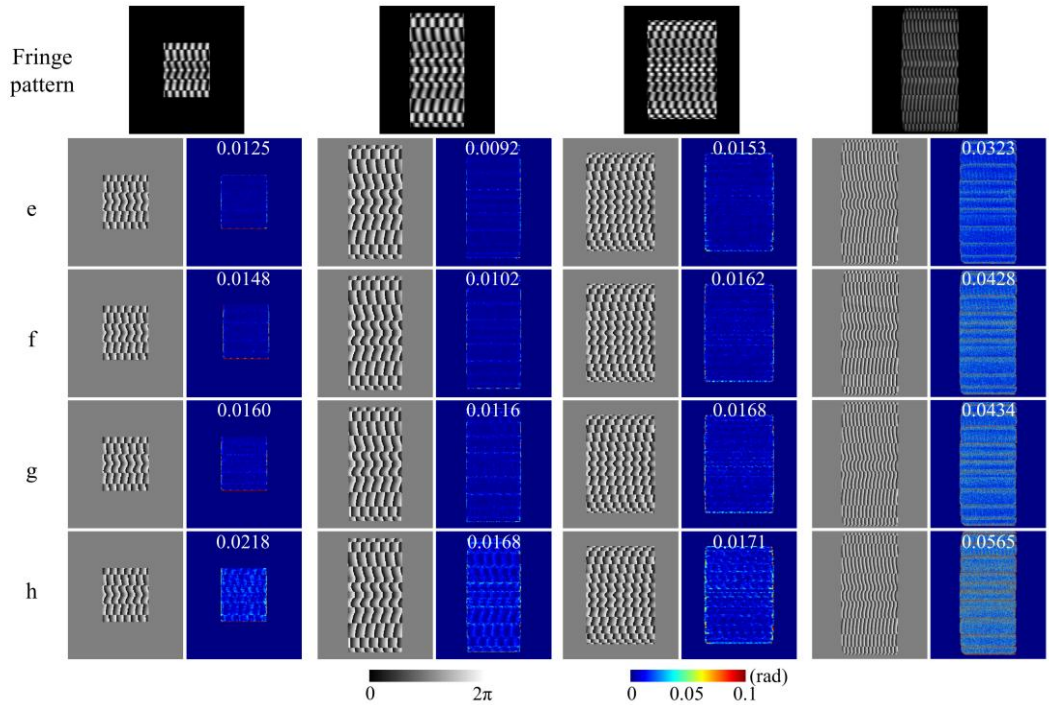

**Fig. 14.** Comparison of wrapped-phase prediction results by four networks under non-uniform motion. Rows e–h corresponds to the four methods (e: CD-DCU-Net, f: TS-U-Net, g: DR-U-Net, h: U-Net), and the topmost row shows the original fringe patterns of different gears. For each gear, the two columns below show the predicted wrapped phase (left) and the error heatmap with respect to the ground truth (right), with RMSE values annotated.

*3.5 Analysis of Loss Functions after Adding Concrete Dropout*

The loss function is improved by introducing a temperature constant into the normal distribution loss and increasing the weight of the mean squared error (MSE) term. This enables the model to focus more on prediction accuracy and to flexibly balance the contributions of different loss terms during training. As shown in Fig. 14, after adding the temperature constant, the MSE loss decreases by about 40% compared to the version without the temperature constant, directly reflecting the effect of loss optimization. Furthermore, the prediction errors of the wrapped phase under both uniform and non-uniform motion are compared for models with and without the temperature constant. The error distributions are shown in Figs. 16 and 17, the results demonstrate that introducing the temperature constant and increasing the MSE weight improves the prediction accuracy of the wrapped phase by 10%–20% compared to the model without the temperature constant.

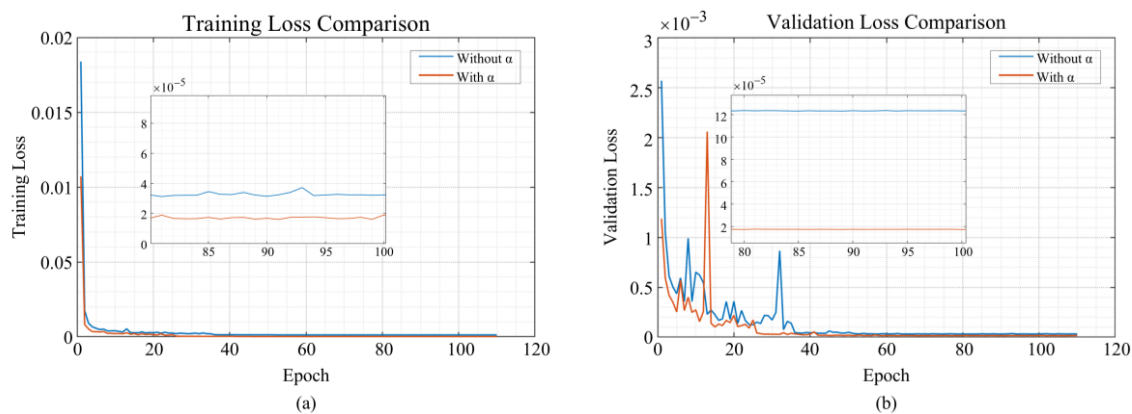

Fig. 15. Comparison of training processes before and after introducing the temperature constant(a) Training set loss curves and (b) validation set loss curves. It can be observed from the figures that the training process after introducing the temperature constant (red line) has better convergence performance compared with the original method (blue line). Especially on the validation set, the model with the temperature constant shows a more stable convergence trend, with smaller fluctuations in loss values and maintaining a lower level in the later training stage (as shown in the enlarged part of the figure).

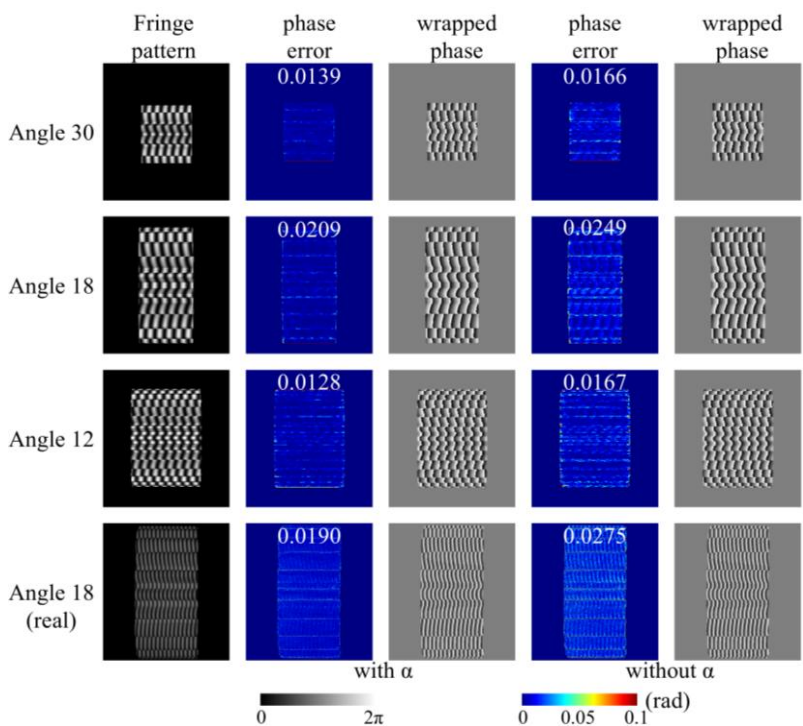

Fig. 16 Comparison of wrapped phase prediction with and without the temperature constant α for four gear samples (pitch angles of 30°, 18°, 12°, and real gear with 18°) under uniform rotation. The first column shows the fringe patterns. The next four columns are grouped in pairs:

the first group (with α) displays the wrapped phase RMSE and wrapped phase predicted by the model with the temperature constant; the second group (without α) shows the corresponding results for the model without the temperature constant.

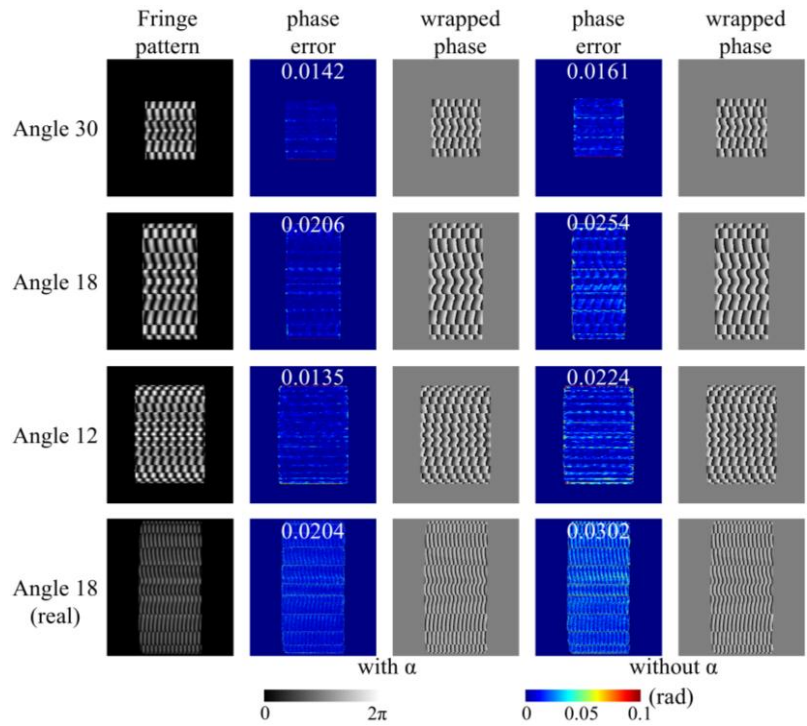

**Fig. 17.** Comparison of wrapped phase prediction with and without the temperature constant α for four gear samples (pitch angles of 30°, 18°, 12°, and real gear with 18°) under non-uniform rotation. The first column shows the fringe patterns. The next four columns are grouped in pairs: the first group (with α) displays the wrapped phase RMSE and wrapped phase predicted by the model with the temperature constant; the second group (without α) shows the corresponding results for the model without the temperature constant.

*3.6 Uncertainty Analysis*

In stripe projection measurement, the reliability assessment of prediction results is of vital importance. The method proposed in this paper not only outputs the predicted values of sine term (M) and cosine term (D), but also obtains their respective variance estimates through probabilistic modeling, thereby enabling quantification of the uncertainty of prediction results. Such uncertainty estimation is of great significance for evaluating the credibility of measurement results, as it helps identify regions with high prediction uncertainty and provides a basis for subsequent measurement optimization.

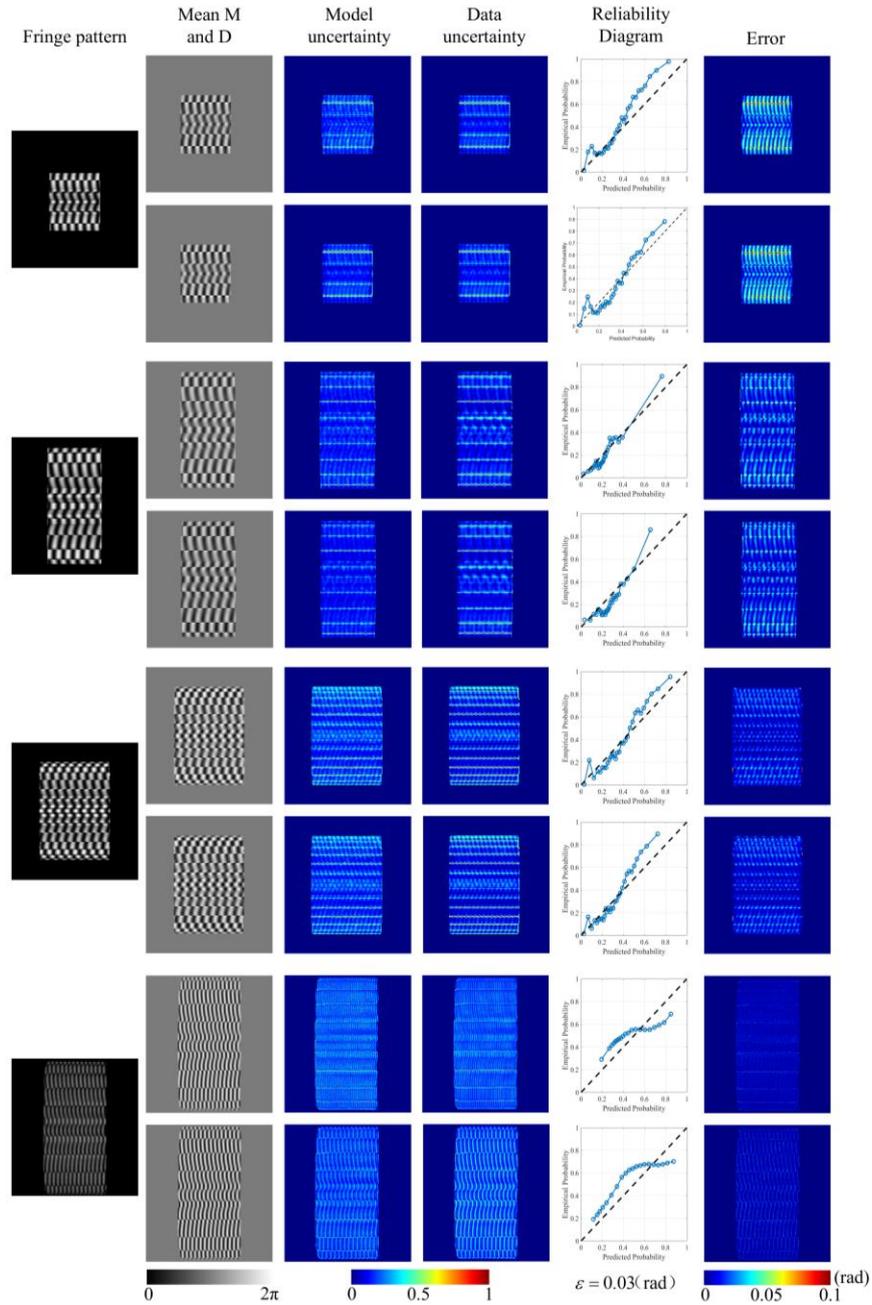

**Fig. 18.** presents the calibration results for four gears in the validation set. For each gear, from left to right are shown: the initial fringe pattern input image, predicted mean values of the sine term (M) and cosine term (D), model uncertainty, data uncertainty, reliability diagram, and actual prediction error. For each gear, the top and bottom rows display the prediction results for M and D, respectively. In the reliability diagrams, the credible interval width ε was set to 0.03 rad, and the predicted results (blue points) lie close to the diagonal line (black dashed line), indicating good agreement between the model's uncertainty estimation and the actual errors.

To evaluate the accuracy of the model's uncertainty prediction, a reliability diagram[51] was employed for calibration analysis. Prediction samples in the validation set were divided into M=30 intervals according to the magnitude of uncertainty, with the credible interval width ε set to 0.03 rad. The average confidence and empirical accuracy were then computed for each interval to assess the calibration performance. Ideally, if the uncertainty estimation is accurate, the average confidence should closely match the empirical accuracy, and the reliability diagram should align with the diagonal line. As shown in the reliability diagram, although the model demonstrates a slightly underconfident (conservative) tendency, its calibration curve is generally close to the diagonal, indicating good overall calibration. Moreover, the uncertainty predicted by the model effectively reflects the actual errors, as illustrated in Figure 18, which presents the predicted values, actual errors, and reliability diagrams of M and D for four gears in the validation set.

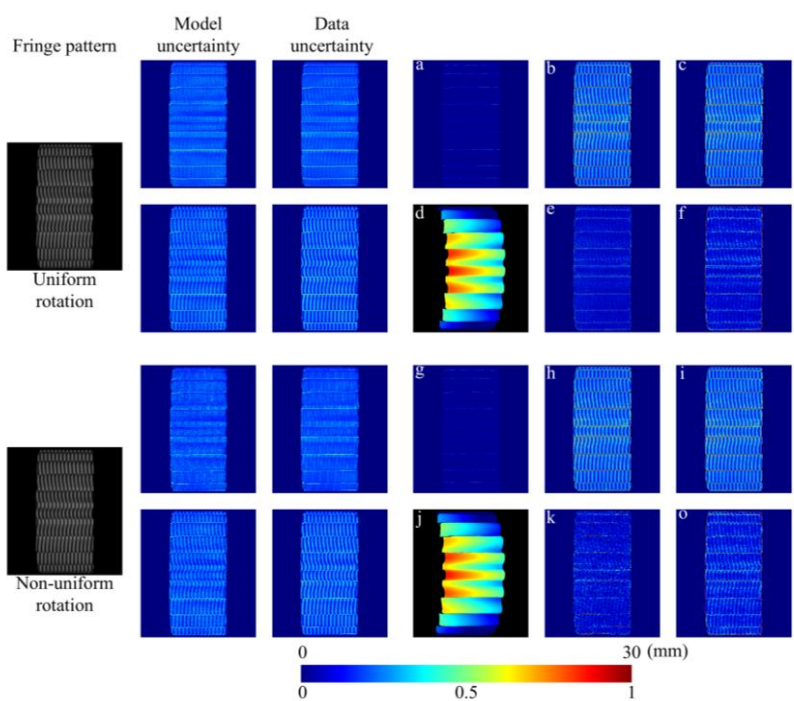

**Fig. 19.** Uncertainty and error analysis for an actual gear under uniform and non-uniform rotation. The first two rows correspond to uniform rotation, and the last two rows to non-uniform rotation. The leftmost three columns show the fringe pattern, model uncertainty map, and data uncertainty map (top row for M, bottom row for D). Panels a–c depict the wrapped-phase model uncertainty, wrapped-phase data uncertainty, and combined total uncertainty; d–f show the 3D reconstruction, phase-error map, and point-cloud error; panels g–o follow the same layout for the remaining results. In uniform rotation, the angular increment from frame 1 to 2 and frame 2 to 3 is 0.6°; in non-uniform rotation, the increment is 0.1° from frame 1 to 2 and 0.9° from frame 2 to 3.

*3.7 CD-DCU-Net vs. 3-Step PS and Generalization Testing*

The proposed CD-DCU-Net supports two prediction modes: single deterministic inference, in which Dropout is disabled and a single forward pass using the trained weights is performed, and Monte Carlo sampling, where Concrete Dropout remains enabled and multiple stochastic forward passes are averaged. As shown in Fig. 20, Monte Carlo sampling achieves slightly higher wrapped-phase prediction accuracy than single deterministic inference, particularly on noisy real-world data where its inherent denoising effect is beneficial. Fig. 21 presents a direct comparison of 3D reconstruction results between the two methods, with Monte Carlo sampling yielding marginally improved surface fidelity. While Monte Carlo sampling provides a modest accuracy gain, single deterministic inference remains preferable in scenarios with strict computational or real-time constraints due to its minimal overhead.

In addition, to further evaluate the generalization capability of the proposed network, we selected gears from the training set and collected new three-step phase-shifting patterns under frame interval conditions that were not present in the training or validation sets. Specifically, the generalization tests included two types of frame interval settings: (1) uniform intervals, with each frame interval set to 0.5°, 1°, and 2°, respectively; and (2) non-uniform intervals, with three groups of combinations: 0.5° and 1°, 1° and 2°, and 2° and 3° (i.e., the intervals between adjacent frames were different). The experimental results show that CD-DCU-Net can still maintain high prediction accuracy under these unseen frame interval conditions, demonstrating strong generalization capability. However, as shown in Fig. 22, to achieve high-precision 3D reconstruction (point cloud error below 0.05 mm), the maximum angular distance between the first and third frames of the three-step phase-shifting input should not exceed the largest rotation angle in the training set (1.7° in this study). When this maximum distance is greater than 2°, the prediction accuracy drops significantly.

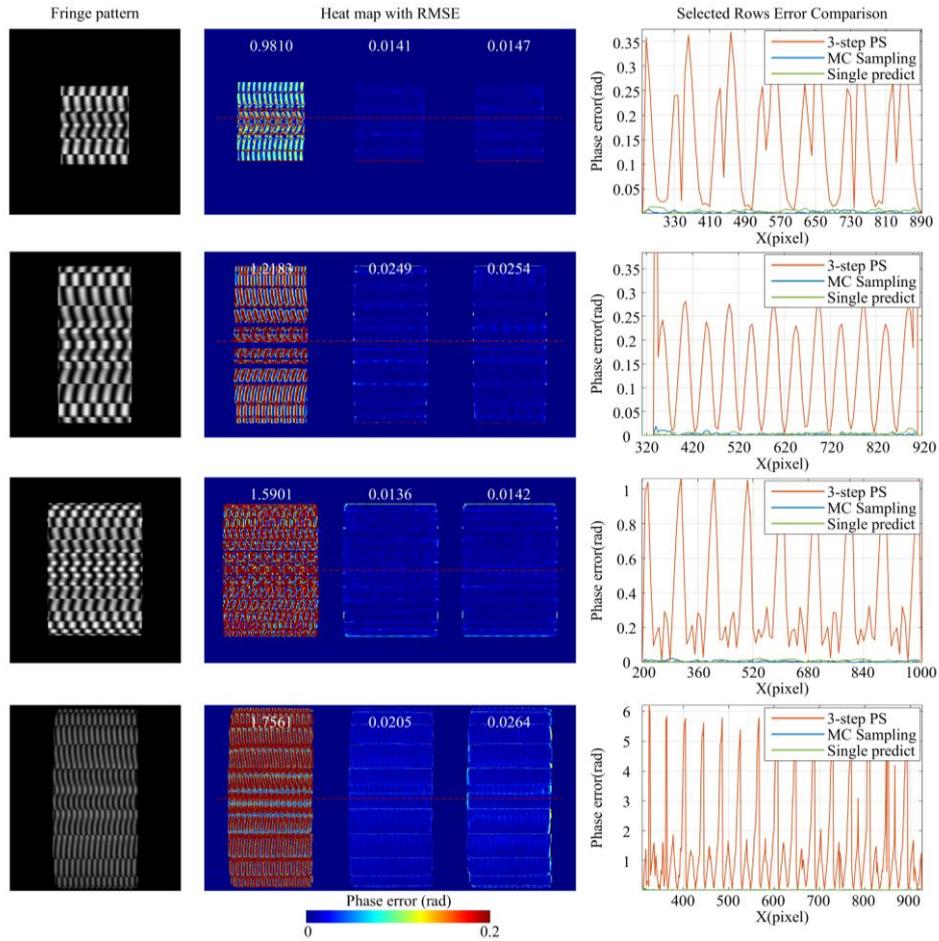

**Fig. 20.** Comparison of phase prediction accuracy by different methods. The left column shows fringe patterns of four gears selected from the validation set; the middle column presents phase-error heatmaps for the traditional three-step phase-shifting method, Monte Carlo sampling prediction, and single deterministic prediction (from left to right), with RMSE values annotated above each heatmap; and the right column displays phase-error curves along the red-dashed rows in the heatmaps (red: three-step PS; blue: MC sampling; green: single prediction). From the RMSE values in the heatmaps, it can be seen that for the four gears, the errors of the traditional 3-step PS method are 0.9810 rad, 1.2183 rad, 1.5901 rad, and 1.7561 rad; the errors of Monte Carlo sampling prediction are reduced to 0.0141 rad, 0.0249 rad, 0.0136 rad, and 0.0205 rad; while single deterministic prediction further lowers the errors to 0.0147 rad, 0.0254 rad, 0.0142 rad, and 0.0264 rad. The phase-error curves further show that both deep-learning methods (Monte Carlo sampling – blue line; single prediction – green line) significantly suppress phase errors compared with the traditional method (red line), with Monte Carlo sampling achieving the highest accuracy.

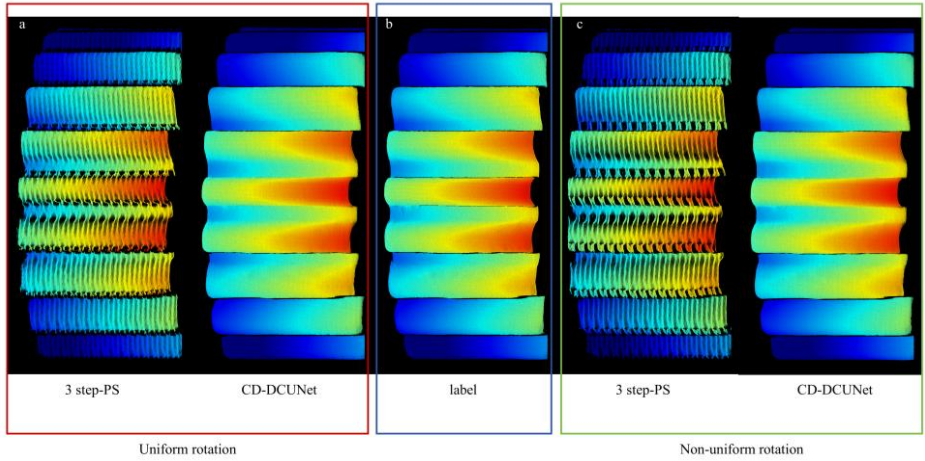

**Fig. 21.** Comparison of 3D reconstruction results of the label, the proposed method, and the three-step phase-shifting method under uniform motion and non-uniform motion. a, b, and c represent the reconstruction results of uniform motion, the label value (18 step-PS), and the reconstruction results of non-uniform motion, respectively. In uniform rotation, the angular increment is 0.6° from frame 1 to 2 and 0.6° from 2 to 3; in non-uniform rotation, it is 0.2° from frame 1 to 2 and 1.5° from 2 to 3.

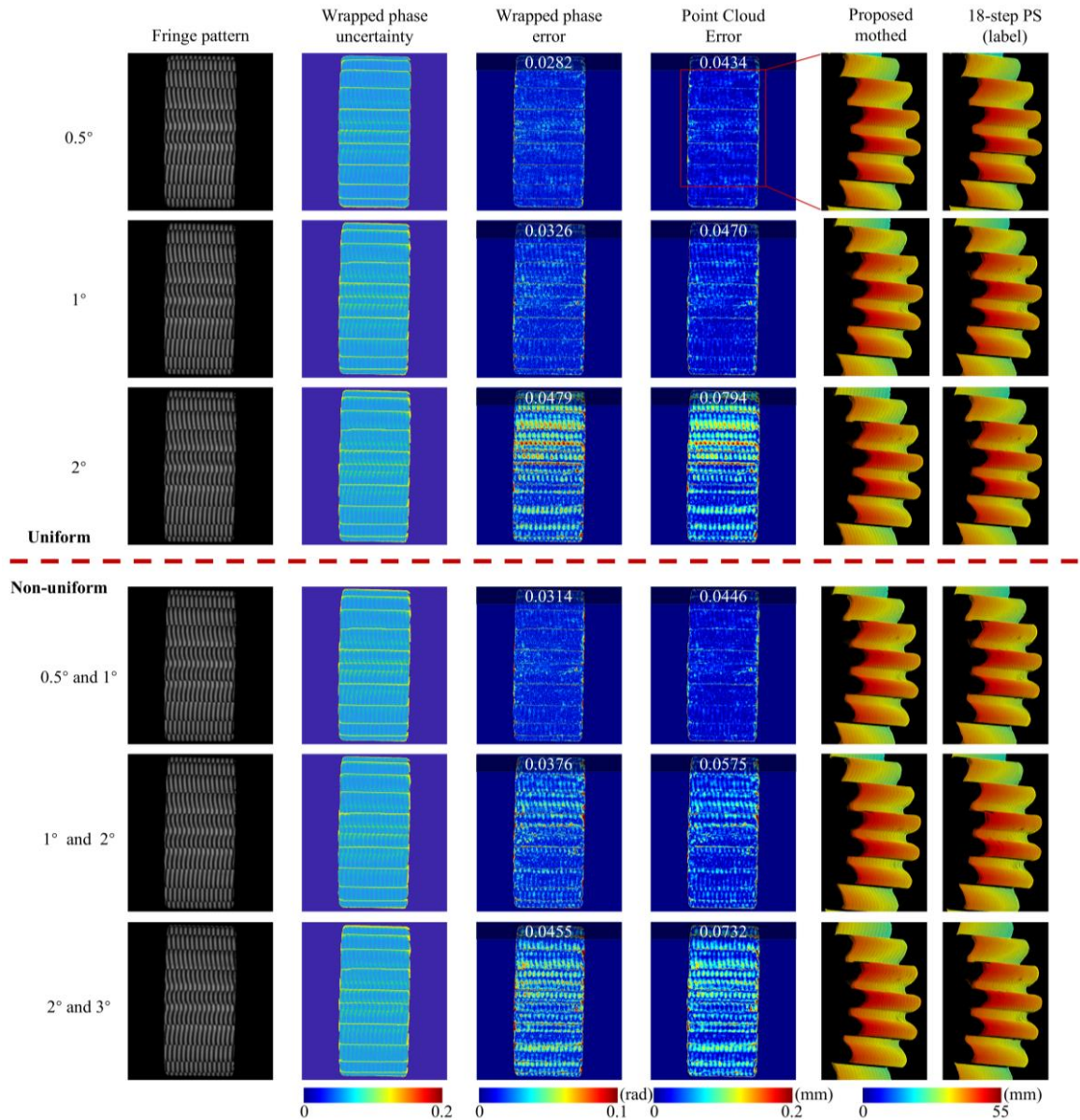

**Fig. 22.** Performance of CD-DCU-Net on the test set under different motion types. The upper part shows results for uniform motion (frame intervals of 0.5°, 1°, and 2°), while the lower part presents non-uniform motion (combinations such as 0.5°, 1°, and 2°). Each row corresponds to a specific frame interval setting, and the columns from left to right are: fringe pattern, uncertainty, wrapped phase error, point cloud error (with the red box indicating the 3D reconstruction range), and the 3D results from the proposed and 18-step PS methods.

## 4. Conclusion

The CD-DCU-Net proposed in this study combines Concrete Dropout and dual 3×3 convolutions in the output layer to achieve joint prediction of wrapped phase and pixel-wise uncertainty. The network is pretrained with a small amount of simulated data generated by the Unity-based digital twin system and then further trained on the target dataset, thus realizing efficient transfer learning. Meanwhile, the introduced weighted loss function effectively balances phase prediction and uncertainty estimation. Experimental results show that the proposed method outperforms the traditional 3-step PS method in measurement accuracy, dynamic error correction, and uncertainty estimation, making it suitable for dynamic gear measurement scenarios.